\documentclass[prd,nofootinbib,preprint,superscriptaddress]{revtex4-1}
\usepackage{amsmath}
\usepackage{graphics}
\usepackage{epstopdf}
\usepackage{graphicx}
\usepackage{subfigure}

\makeatletter
\renewcommand{\fnum@table}{\textbf{\tablename~\thetable}}
\renewcommand{\fnum@figure}{\textbf{\figurename~\thefigure}}
\makeatother

\newcommand {\be}{\begin{equation}}
\newcommand {\ee}{\end{equation}}
\newcommand {\ba}{\begin{eqnarray}}
\newcommand {\ea}{\end{eqnarray}}

\begin{document}

%%%%%%%%%%%%%%%%%%
%%  TITLE PAGE  %%
%%%%%%%%%%%%%%%%%%

\vspace*{10mm}

\title{Sensitivities to secret neutrino interaction at FASER$\nu$ \vspace*{1.cm} }

\author{\bf  Majid Bahraminasr }
\affiliation{Institute for
	research in fundamental sciences (IPM), PO Box 19395-5531, Tehran,
	Iran}
\author{\bf Pouya Bakhti}
\affiliation{Institute for
	research in fundamental sciences (IPM), PO Box 19395-5531, Tehran,
	Iran}
\affiliation{Jeonbuk National University, Jeonrabuk-do 54896, South Korea}
\author{\bf  Meshkat Rajaee}
\affiliation{Institute for
	research in fundamental sciences (IPM), PO Box 19395-5531, Tehran,
	Iran}
\affiliation{Jeonbuk National University, Jeonrabuk-do 54896, South Korea}
\begin{abstract}
  \vspace*{.5cm}
  
We study the  impact of the coupling of neutrinos with a new light neutral gauge boson, $Z^\prime$, with a mass of less than 500 {\rm MeV} in FASER$\nu$  experiment.
Scenarios in which a light gauge boson is coupled to neutrinos are motivated within numerous
contexts which are designed to explain various anomalies in particle physics and cosmology. This interaction  leads to a new decay mode for charged mesons to a light lepton plus neutrino and $Z^{\prime}$, ($\pi^+(K^+)\to e^+ \nu Z^\prime$) followed by the   subsequent decay of   $Z^\prime$    into the pair of neutrino and anti-neutrino, ($Z^\prime \to \nu\bar{\nu}$).
FASER$\nu$, the Forward
Search Experiment at the LHC, has the potential to detect collider neutrinos for the first time. In particular, the FASER$\nu$ emulsion detector will provide the opportunity to detect $\tau$-neutrinos and to measure their energies. Using this ability of FASER$\nu$ emulsion detector, we investigate the potential of  FASER$\nu$ experiment and the proposed upgraded version of this experiment, FASER2$\nu$, to constrain the coupling of a neutrino with the light gauge boson.

\end{abstract}

\maketitle

%%%%%%%%%%%%%%%%%%%%%%%%%%%%%%%%%%%%
\section{Introduction}
%%%%%%%%%%%%%%%%%%%%%%%%%%%%

Although particle colliders
produce neutrinos of all flavors copiously, collider neutrinos have not yet been detected for two main reasons. First, neutrinos interact very weakly and second, collider detectors miss the enormous flux of high-energy neutrinos streaming down the beam pipe and are blind to the regions along the beamline.
FASER, the Forward Search Experiment at the Large Hadron Collider (LHC) is going to be located
480 m downstream of the ATLAS interaction point along the beam collision axis. FASER's location is ideal to cover this blind region.
Having such an ideal location,
FASER will provide sensitive searches for light and weakly interacting particles
in Run 3 from
2022 to 2024. FASER’s neutrino detection capability is briefly discussed in \cite{Ariga:2018zuc} and more detailed studies on the
detector design are reported in \cite{ariga}.

FASER$\nu$ is a sub-detector that will be able to detect the collider neutrinos for the first time \cite{Abreu:2019yak}. FASER$\nu$ will be located in front of FASER spectrometer at CERN \cite{Abreu:2019yak}.
Depending on the neutrino flavor, the mean energy of neutrinos interacting in FASER$\nu$ is between 600~{\rm GeV} and 1~{\rm TeV} with a significant number of neutrino events up to 3 {\rm TeV}. FASER$\nu$ will detect the most energetic neutrinos from the known source.
%, although, the most energetic neutrinos ever.
The other advantage of FASER$\nu$ is its emulsion detector that has the greatest precision to detect tau-neutrinos.
Detecting a tau-neutrino requires that the neutrino beam has enough energy to produce a tau lepton $(E_{\nu_\tau} > 3.5 ~{ \rm GeV})$. On the other hand, tau leptons are short-lived and decay promptly. This makes their identification extremely hard.
Having an integrated luminosity of 150 ${\rm fb^{-1}}$, FASER$\nu$ will be a good apparatus to identify the tau-neutrinos and to
study the new neutrinophilic interaction with tau-neutrino detection.

Being able to observe these interactions and reconstruct their energies, FASER$\nu$ will probe the production, propagation, and interactions of neutrinos at very high energies ({\rm TeV}).
Since FASER$\nu$ benefits from high-energy neutrinos producing from very high-energy mesons, it is interesting to investigate the possibility of interaction of neutrinos with new light particles using this experiment. FASER2 is a proposed upgraded version of the FASER experiment, with several hundred to thousand times larger data collection due to higher luminosity and spectrometer dimensions. It is also proposed to use a larger emulsion neutrino detector in front of the FASER2 tunnel.

In the present paper, we study the scenario in which neutrinos couple to a light gauge boson $Z^\prime$ with a mass smaller
than $ \sim 500$ {\rm MeV}.
Neutrinophilic new gauge interaction with a light gauge boson is motivated by so-called $\nu$-DM models. As proposed in \cite{Chu:2015ipa, Hooper:2007tu, Aarssen:2012fx},
they can help to solve small-scale structure problems that appear in the canonic collisionless cold dark matter paradigm. If the new gauge boson couples to matter fields, scattering experiments will be able to detect it. Besides, it can affect the neutrino oscillation in the matter. However, if $Z^\prime$ couples only to the neutrinos, it will not
affect neutrino oscillation in the matter or elastic scattering of neutrinos off nuclei. Moreover, assuming that only neutrino couples to the new gauge boson, $Z^\prime$
decays only into neutrinos at tree level, appearing as missing
energy in the experiments.
Assuming the standard meson two-body decay ($M \rightarrow l\nu$), the decay rate is suppressed by $ m^2 _l /m^2 _M $.
Assuming the three-body decay of the Meson ($M \rightarrow l\nu Z^\prime $), the decay rate receives an enhancement of $ m^2 _M /m^2 _{Z^\prime} $ from longitudinal polarization; Thus, it can provide us an opportunity to search for even small gauge
coupling.

In this work, we study how decays of charged mesons ($M \rightarrow l\nu Z^\prime $) can provide us information about neutrinos interacting with new light particles. We will investigate the sensitivity of leptonic decay of charged mesons to the interaction of
neutrinos with $Z^\prime$. In our scenario, a new gauge boson with a mass of less than 500 {\rm MeV}
can be produced via $M \rightarrow l \nu Z^\prime $ and subsequently $Z^\prime $ decays into a pair of a neutrino and an anti-neutrino before reaching the near detector. The
produced neutrinos can be detected, providing us with information on the intermediate $Z^\prime$.

Several models for neutrino interaction with the new light gauge boson have been proposed in the literature. One possible underlying model that can lead to this interaction is proposed in reference \cite{farzan,farzaan}, introducing a new fermion of a mass of the order of {\rm GeV} that is charged under an extra U(1) gauge symmetry. This new fermion is mixed with neutrino and the active neutrinos will obtain interactions of this form.

This interaction can lead to a new mode of meson decay, $M\to l\nu Z^\prime$, and subsequently, $Z^\prime$ decays to neutrino anti-neutrino pair. Charged meson decays and short-baseline accelerator-based neutrino experiments are sensitive probes of this neutrinophilic interaction of light new particles.
%Meson decay experiments can only constrain $ \sqrt{\sum_\alpha |g_{e\alpha}|^2}$.
The most stringent current constraint on the coupling comes from kaon decay rare event measurement at NA62 in the range of {\rm MeV} to a few ten {\rm MeV} \cite{Lazzeroni:2012cx, Bakhti:2017jhm}. Furthermore, the near detector of DUNE \cite{Acciarri:2015uup}, will constrain the scenario more stringently, due to a large number of statistics and tau-neutrino detection with low background \cite{Bakhti:2018avv}.
FASER$\nu$ has the advantage of producing neutrinos from massive mesons such as charm mesons and therefore, it can help to constrain the coupling in larger mass range of $m_{Z'}$. Moreover, it benefits from large efficiency of tau neutrinos as well as large values of neutrino energy, so it has a great potential to determine $g_{ee} $, $g_{e \tau }$, $g_{e\mu}$ and $g_{\mu\tau} $
separately by studying the electron neutrino and tau neutrino signals. In this paper we study this possibility to constrain $g_{ee} $, $g_{e \tau }$, $g_{e\mu}$ and $g_{\mu\tau} $ using FASER$\nu$ and FASER2$\nu$ data. %In this paper, we are mainly interested in the $g_{ee}$ and $g_{e \tau}$ since they can lead to kinematically favored decay modes $ \pi^+ \rightarrow e^+ \nu_e Z^{\prime} $ and $ \pi^+ \rightarrow e^+ \nu_\tau Z^{\prime} $

%In this work, we will investigate the discovery potential of FASER$\nu$ and FASER2$\nu$ for non-zero $g_{ee}$ , $g_{e \tau}$, $g_{\mu\tau}$ and $g_{e \mu}$. We will discuss how the produced tau-neutrinos and electron-neutrinos detected by FASER$\nu$ can provide information about the flavor structure of $Z^{\prime}$ coupling.

The present paper is organized as follows. In Sec.~II, we present a short review of the new Lagrangian and the decay rates. In Sec.~III, the details of the FASER$\nu$ experiment and our simulation are discussed. In Sec.~IV, we present our results. Sec.~V is dedicated to the summary and discussion.

%%%%%%%%%%%%%%%%%%%%%%%%%%%%
\section{Leptophilic gauge interaction, meson decay, and neutrino \label{theo}}
%%%%%%%%%%%%%%%%%%%%%%%%%%%%%%%%%%%%%%%

The interaction of the neutrino of flavor $\alpha$ with the new vector boson $Z^\prime$ is given by

\be \sum_{\alpha ,\beta} g_{\alpha \beta}Z'_\mu\bar{\nu}_\alpha \gamma^\mu \nu_\beta \ee
where $g_{\alpha \beta}$ is the the couplings between the new
light boson $Z^{\prime}$ and neutrinos of flavor $\alpha$ and $\beta$.
There are various underlying models leading to such an interaction. Notice that this secret interaction of neutrino with new gauge boson suffers from non-invariance under $SU(2)_L$.
Gauging anomaly free combination of lepton flavors and baryon number can lead to this interaction. Gauging various combinations of lepton flavors and baryon
number $a_e L_e + a_\mu L_\mu + a_\tau L_\tau +b B$ can lead to such an interaction, where $a_e, a_\mu, a_\tau$ and $ b$ are real numbers satisfying the anomaly cancellation condition. There are strong bounds on the coupling of the
electron to $Z^\prime$ from various observations \cite{Kling:2020iar}.
Current constraints on their parameter spaces and
the sensitivity of DONuT and as well as the future emulsion detector experiments FASER$\nu$,
LHC and SHiP on four scenarios of anomaly free U(1) gauge groups corresponding to the $B-L$, $B-L_\mu - 2L_\tau$,
$B-L_e -2 L_\tau$ and $B -3 L_\tau$ are presented in \cite{Kling:2020iar}. The strongest direct constraints in parts of the parameter
space of the $B - L_e - 2 L_\tau$ and $ B - 3 L_\tau$ models are imposed by the DONuT experiment. Since there are strong
bounds on the coupling of the electron to $Z'$, we will not focus on this class of models.

Moreover, as it is discussed in ref. \cite{farzan}, another possibility is introducing a new Dirac fermion which is charged under the new U(1) gauge symmetry and can mix with the active neutrinos. Let us briefly review this scenario. The new fermion $\Psi$ is assumed to be charged under the new $U(1)$ and can mix with $\nu_\alpha$. Let us denote the gauge coupling by $g_\Psi$, the gauge interaction term can be written as $g_\Psi Z'_\mu \bar{\Psi} \gamma^\mu \Psi$. The active neutrinos of flavor $\nu_\alpha$ can be written as a linear combination of mass eigenstates $\nu_i$ as $ \label{nuA}\nu_\alpha =\sum_{i=1}^4 U_{\alpha i} \nu_i$, where $\nu_4$ is the heavier state giving the main contribution to $\Psi$. We assume $\nu_4$ to be heavier than the charged meson $M^+$, therefore, in the decay $M^+ \to l_\alpha^+ \nu+X$, where $X$ could be any state, the coherent $\nu$ state is not exactly equal to $\nu_\alpha$ and is a linear combination of $\nu_1$, $\nu_2$ and $\nu_3$ which cannot be perpendicular to $\Psi$. Integrating out the heavy fourth state, the light active neutrinos receive a coupling of the form $ g_{\alpha\beta } Z_\mu' \bar{\nu}_\alpha \gamma^\mu \nu_\beta $ in which
$ g_{\alpha\beta }=g_\Psi|U_{\Psi 4}|^2 U_{\alpha 4}U_{\beta 4}^* \simeq g_\Psi U_{\alpha 4}U_{\beta 4}^*. $ As a result, three-body decays $M \to l_{\alpha} \nu_\beta Z'$ can take place with a rate proportional to $|g_{\alpha \beta}|^2$. $Z'$ will subsequently decay into $\bar{\nu}_\alpha \nu_\beta$ again with a rate proportional to $|g_{\alpha \beta}|^2$.
However, if $\nu_4$ is lighter than the parent charged lepton, we cannot integrate it out and the picture will be different and this is not the case we are interested in this work.
One method to mix $\Psi$ with $\nu_\alpha$ is to introduce a new Higgs doublet $H^\prime$ charged under the new $U(1)$ such that its vacuum expectation value induces a mixing between $\Psi$ and $\nu_\alpha$ via a Yukawa coupling of the form $ \bar{L}_\alpha H^{\prime T}c\Psi$ \cite{Farzan:2016wym}. As discussed in this reference, in this scenario, $\Psi$ cannot be lighter than a few MeV, otherwise, it contributes as an extra relativistic degree of freedom in the early Universe. On the other hand, it cannot be heavier than a few GeV to maintain the perturbative region and to satisfy the unitarity bounds. In the other model described in detail in \cite{farzaan} a neutral Dirac $N$ and a new scalar singlet $S$ charged under $U(1)$ are introduced with interaction terms similar to that in the inverse seesaw mechanism: $Y_\alpha \bar{N}_RH^TcL_\alpha+\lambda_LS \bar{\Psi}_RN_L$. and $U_{\alpha 4}$ is given by $Y_\alpha \langle H \rangle \lambda_L \langle S\rangle/( m_Nm_\Psi)$.
In this class of models, only neutrinos couple to $Z'$ at tree level. Thus, they escape from the bounds on the coupling of the corresponding charged leptons to $Z'$. %We are only interested in the $g_{ee}$ and $g_{e\tau}$ elements because they can lead to kinematically favored decay modes $\pi^+ \to e^+ \nu_e Z'$ and $\pi^+ \to e^+ \nu_\tau Z'$, respectively.
The bounds on
the deviation of the PMNS mixing matrix from the unitarity can be translated into the bounds on $g_{\alpha \beta}$ \cite{Farzan:2016wym}.
For gauge coupling in the perturbative range, $g_\Psi\stackrel{<}{\sim}4$, the bound from unitarity which is $|U_{e4}|^2<2.5 \times 10^{-3}$ \cite{Fernandez-Martinez:2016lgt} can be translated as $g_{ee} \stackrel{<}{\sim} 10^{-2}$. Notice that in this case, we are not introducing a new source of lepton flavor violating (LFV) so no strong bound comes from $\mu \to e \gamma$ and from similar LFV processes. The unitarity bound on $|U_{e4}U_{\tau 4}^*|$ is $3.7 \times 10^{-3}$ \cite{Fernandez-Martinez:2016lgt} can lead to $g_{e\tau}\stackrel{<}{\sim}10^{-2}$.
The LFV process $\tau \to e \gamma$ does not yield a strong bound since it is GIM suppressed \cite{Farzan:2016wym}.

Let us now compute the flux of neutrinos from meson decay as well as from subsequent $Z^\prime$ decay.
The new interaction leads to a new decay mode of meson decay to electron, neutrino and $Z^\prime$, with the decay rate of \cite{Bakhti:2017jhm}
\begin{equation}\label{decayrate}
\Gamma(M\longrightarrow l_\alpha\nu Z')=\frac{1}{64\pi^3m_M}\int_{E_l^{min}}^{E_l^{max}}\int_{E_\nu^{min}}^{E_\nu^{max}} dE_l dE_\nu \sum_{spins}\vert {\cal M} \vert^2.
\end{equation}

Neglecting the neutrino and lepton masses, the amplitude is
\begin{align}
\label{Msquared2}\sum_{spins}\vert {\cal M} \vert^2 &= (\sum_\beta g_{\alpha \beta}^2)G_F^2f_M^2V_{qq'}^2 \left( m_M^2+m_{Z'}^2-2m_ME_{Z'}\right.\\
& \left.+\frac{(m_M^2-m_{Z'}^2-2m_ME_l)(m_M^2-m_{Z'}^2-2m_ME_\nu)}{m_{Z'}^2}\right),
\end{align}
where $G_F$ is the Fermi constant, $V_{qq^\prime} $ and $f_M$ are the relevant CKM mixing element and meson decay constant, respectively. In the case of $m_e$, integration limits are given by
$$E_e^{min}=m_e,~~~~~~~~~~~~~~~~~~~~~~~~~~~~~~~~~~~~ E_e^{max}=\frac{m_M^2-m_{Z'}^2}{2m_M},$$
$$E_\nu^{min}=\frac{m_M^2-m_{Z'}^2-2m_ME_e}{2m_M},~~~~~~~~~ E_\nu^{max}=\frac{m_M^2-m_{Z'}^2-2m_KE_e}{2(m_M-2E_e)}.$$
Notice that the above formulas are valid if we can neglect the lepton mass. In this case the decay rate can be calculated analytically. In the case of meson decay to muon ($M \rightarrow \mu \nu Z^\prime$) and tau ($M \rightarrow \tau \nu Z^\prime$), we calculate the meson decay rate numerically. %$\Gamma _{\pi \rightarrow l \nu Z^\prime} / \Gamma_{SM}(M \rightarrow l \nu)$ is indicated in Fig.~\ref{gammamuon}. To plot this figure, we have assumed $g_{\alpha \beta} = 1$.
%\begin{figure}
%\begin{center}
%\subfigure[]{\includegraphics[width=0.49\textwidth]{Rpi.pdf}}
%\subfigure[]{\includegraphics[width=0.49\textwidth]{RK.pdf}}
%\subfigure[]{\includegraphics[width=0.49\textwidth]{RDs.pdf}}
%\end{center}
%\caption[]{ $\Gamma _{M \rightarrow l \nu Z^\prime} / \Gamma_{SM}(M \rightarrow l \nu)$ as a function of $M_{Z^\prime}$ plotted for pion, kaon and $D_s$ decay channels. }
%\label{gammamuon}
%\end{figure}

The number of neutrinos coming from $Z^\prime$ particles decaying before reaching the detector is given by
\begin{equation}\label{Eq.NOFzpdecay}
N=N_0\left(1-e^{-\Gamma L / \gamma} \right)
\end{equation}
where $N_0$ is the number of produced $Z^\prime$, $L$ is the distance between the production point of $Z^\prime$ and the detector and $\gamma=E_{Z^\prime}/m_{Z^\prime}$ is the boost factor. As it can be seen from equation \ref{Eq.NOFzpdecay}, if $\Gamma L/ \gamma \gg1$ almost all of the $Z^\prime$ particles decay before reaching the detector.

The estimated number of neutrinos coming from different meson decays passing through
FASER$\nu$ detector is given in ref \cite{ariga}, assuming an integrated luminosity of 150 ${ \rm fb ^{-1}}$
for Run 3 at the 14{\rm TeV} LHC. Considering the spectrum of neutrinos coming from pion, kaon and charm meson decay, we can reconstruct the spectrum of the initial meson. In the rest frame of the meson, the total spectrum of the neutrino produced from both meson and $Z'$ decay is given by

\begin{equation}\label{Eq.anuspec}
(\frac{dN_\nu}{dE_\nu})_{r.o.M}= (\frac{dN_\nu}{dE_\nu })_{r.o.M}^{Z' decay}
+ \frac{N_0}{ \Gamma (M\longrightarrow l\nu Z')}\frac{d\Gamma ( M\longrightarrow l\nu Z')}{dE_{\nu}}
\end{equation}
where $N_0$ is the total number of the neutrinos produced from meson decay. $ \Gamma (M\longrightarrow e\nu Z')$ and $\frac{d\Gamma ( M\longrightarrow e\nu Z')}{dE_{\nu}}$ is determined from Eq.~\ref{decayrate}. The spectrum of neutrinos produced from $Z^\prime$ decay is determined with integration over the $Z^\prime$ spectrum in rest frame of meson multiplied by neutrino spectrum for a specific energy of $Z^\prime$ as follows
\be
(\frac{dN_\nu}{dE_\nu })_{r.M}^{Z'~ decay}= \sum_i\int_{E_{Z'}^{min}}^{E_{Z'}^{max}} dE_{Z'} \frac{dN_{Z'}}{dE_{Z'}}\mid_{i} (\frac{dN_\nu}{dE_\nu})_{r.M}\mid_{i}
\ee
where $E_{Z'}^{min}=E_\nu+m_{Z'}^2/(4E_\nu)$, $E_{Z'}^{max}=(m_M^2+m_{Z'}^2)/(2m_M)$ and $i$ refers to $Z'$ different polarizations. The neutrino energy at the rest frame of $Z^\prime$ is given by $m_{Z^\prime}/2$, and neutrino spectrum for a specific energy of $Z^\prime$ is calculated with boost of $Z^\prime$. The neutrino spectrum in the lab frame is given by
\begin{equation}\label{Eq.nuflux}
\phi(E_\nu)= \frac{1}{4 \pi L^2} \int_{E_M^{min}}^{E_M^{max}}dE_{M} P_M(E_M) (\frac{dN_\nu}{dE_\nu})_{lab} \frac{d \Omega_{r.M}}{d \Omega_{lab}},
\end{equation}
where $P_M(E_M)$ is the differential meson spectrum in the lab frame and $(\frac{dN_\nu}{dE_\nu})_{lab}$ is the spectrum of the neutrino in the lab frame. $d\Omega_{r.M}/d\Omega_{lab}=(1+v_M)/(4(1-v_M)) \simeq \gamma_M^2$ takes care of focusing of the beam in the direction of the detector. $v_M$ is the meson velocity in the lab frame and $\gamma_M=(1-v_M^2)^{-1/2}$.
For details of the calculation see the Appendix of Ref.~\cite{Bakhti:2018avv} and Ref.~\cite{Bakhti:2017jhm}.

\section{   FASER$\nu$ Experiment And  Light $Z^{\prime}$}

FASER experiment is a spectrometer tunnel with a length of 1.5 m and a radius of 10 cm, 480 m downstream to the ATLAS interaction point. This apparatus will be sensitive to new physics measurements such as dark photons and axion-like particles \cite{Abreu:2019yak}. FASER$\nu$ is a proposal to detect collider neutrinos for the first time using an emulsion detector in front of FASER spectrometer \cite{Abreu:2019yak}. FASER$\nu$ consists of 1.2 tons tungsten plates and 1000 layers of emulsion films. The neutrino beam consists of both neutrino and anti-neutrino, and they will be detected with charged-current deep inelastic scattering from nuclei. The flavor of neutrinos is determined with the charged lepton detection. Assuming standard model cross-section, the total number of 20000, 1300, and 20, respectively for muon, electron, and tau-neutrinos will be detected. The neutrino energy is in the range of a few 10 {\rm GeV} to a few {\rm TeV} with the peak of neutrino interaction between 600 {\rm GeV} to 1 {\rm TeV}. In this energy range, $\nu_\tau$ cross-section is large and approximately equal to electron and muon neutrino cross-sections. The cross-section of deep inelastic scattering is approximately proportional to the neutrino energy.
In Ref. \cite{Abreu:2019yak} the cross-section is calculated by considering NNPDF3.1NNLO parton distribution function \cite{Ball:2017nwa}.

As it is well known, the identification of tau leptons is extremely difficult.
Directly detecting tau-neutrino requires that the neutrino beam has enough
energy to produce a tau particle.  With a spatial resolution of a few ten nm, emulsion detectors are the most sensitive  for detecting the short-lived particles  such as    tau-neutrinos. FASER$\nu$ energy resolution is 30$\%$ and the neutrino energy is determined from leptonic and hadronic energies. The backgrounds of electron neutrino CC interaction detection are the shower of the neutral pion to photon pair, and pion decay to photon and electron-positron pair. The other source of background is muon neutrino CC interaction. Backgrounds of tau-neutrino CC interaction are neutral current interaction and CC interaction of electron and muon neutrino when they are associated with charm production at the interaction vertex. There are also accidental backgrounds. These backgrounds at FASER$\nu$ will be much smaller than other emulsion detectors like OPERA, due to larger neutrino energy. Moreover, by combining FASER and FASER$\nu$, we can distinguish between $\nu_\mu$ and $\bar{\nu}_\mu$. Due to the long lifetime of a muon, the produced muon at FASER$\nu$ will pass through FASER spectrometer.

FASER2 is a proposed experiment similar to FASER, with twenty times larger luminosity and mass detector of 100 times larger than FASER \cite{Abreu:2019yak}. It is also proposed to locate a larger emulsion neutrino detector in front of the FASER2. The mass of the detector is under discussion. Also in Ref. \cite{Abreu:2019yak} a mass of 10 to 1000 tons is proposed while in Ref. \cite{Abreu:2020ddv} a mass of the order of ten tons is mentioned. In our analysis, we have considered the maximum possible future statistics for the detection of neutrinos at FASER2 which is approximately $100$ and $1000$ times larger data than FASER$\nu$. All the details of the analysis of FASER2$\nu$ are the same as FASER$\nu$, except for the statistics. In the following, we will show that FASER$\nu$ has the potential to set a more stringent bound on the coupling for the mass range of $50 ~{\rm MeV} <m_{Z'}< 150~ {\rm MeV}$. However, the upgraded version of FASER$\nu$ will constrain our scenario more stringently than the current bound in the range of $m_{Z'}< 2~ {\rm keV}$ and $3~ {\rm MeV} <m_{Z'}<200~ {\rm MeV}$.

%%%%%%%ffff8%%%%%%%%%%%%%%%%%%%%%%%%%%%%%%%%%%%%%%%%%%%%%%%%%%%%%%%%%%%%%
\begin{figure}
\begin{center}
\subfigure[]{\includegraphics[width=0.49\textwidth]{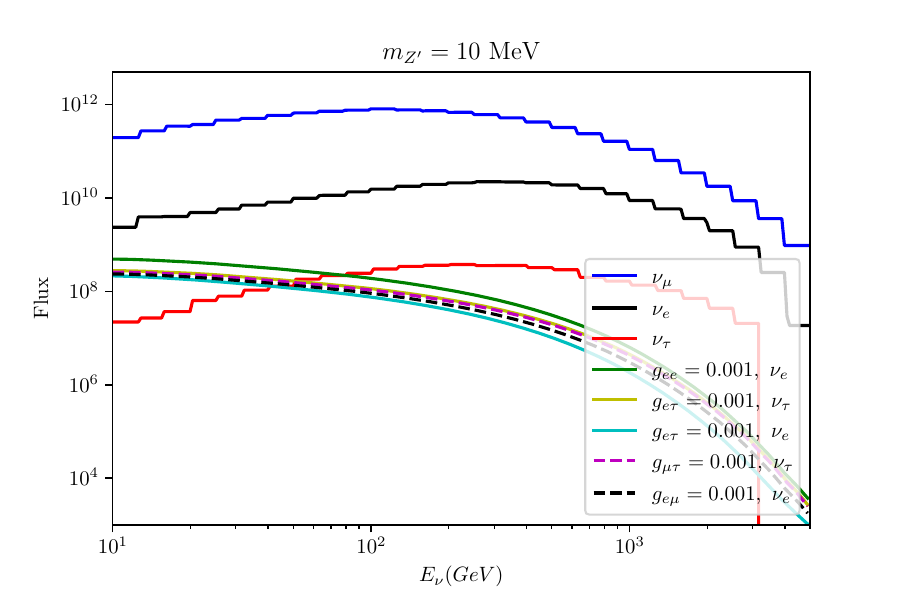}}
\subfigure[]{\includegraphics[width=0.49\textwidth]{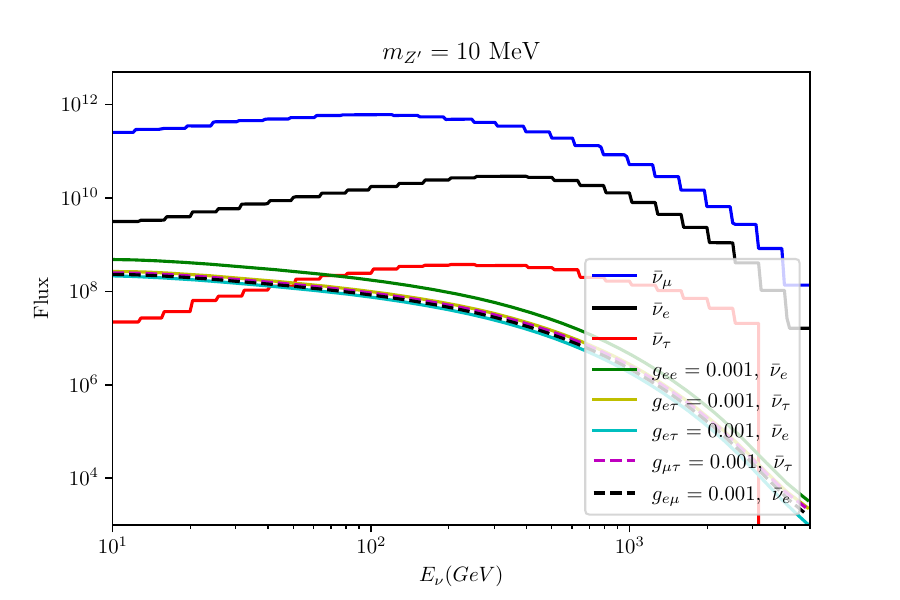}}
\end{center}
\caption[]{The estimated number of  neutrinos (left panel) and antineutrino (right panel) passing through   the detector of FASER$\nu$ experiment for muon, electron, and tau neutrinos assuming an integrated luminosity of   $150~  {\rm fb^{-1}}$ for the Run 3 at the 14 TeV LHC. The blue, black and the red curves are plotted for $\nu_{\mu}$ ($\bar{\nu}_{\mu}$), $\nu_{e}$ ($\bar{\nu}_{e}$) and $\nu_{\tau}$ ($\bar{\nu}_{\tau}$), respectively, in the left (right) panel assuming  the standard model as the true model. To plot other cureves, we have assumed   $m_{Z^\prime}=10~$ {\rm MeV} and $g_{\alpha\beta}$ is set  to  $0.001$.}
\label{flux_nu_anu}
\end{figure}
%%%%%%%ffff8%%%%%%%%%%%%%%%%%%%%%%%%%%%%%%%%%%%%%%%%%%%%%%%%%%%%%%%%%%%%%

\section{ Constraints}

In this section, we discuss how the data from the FASER$\nu$ and FASER2$\nu$ can be used to extract information on the
coupling of neutrinos to the new gauge boson. The spectrum of produced neutrinos from different decay modes is presented in Ref.~\cite{Abreu:2019yak}. $2\times10^{11}$ electron neutrinos, $6\times10^{12}$ muon neutrinos, $4\times10^{9}$ tau-neutrinos and a comparable anti-neutrinos pass through the FASER$\nu$ detector. As mentioned in Ref.~\cite{Abreu:2019yak} tau-neutrinos are mainly produced from $D_s$, strange charm meson (Fig.4 of Ref.~\cite{Abreu:2019yak}). To find the spectrum of $D_s $, we have taken the neutrino spectrum and assumed that all the produced neutrinos are coming from $D_s \rightarrow \tau \nu_{\tau}$. Noticing that the branching ratio of $D_s \rightarrow \tau \nu_{\tau}$ is $(5.6\pm0.4)\%$. As we will see in the following this decay channel is very important for constraining $ g_{\alpha \beta}$, especially for $m_{Z'}>50$ {\rm MeV}.

Considering the new interaction, electron and/or tau-neutrinos can be produced from meson three-body decay, $M\to l\nu_\alpha Z^\prime$, and subsequent $Z^\prime$ decay to neutrino-antineutrino pair.
Pion and kaon leptonic decays are the dominant modes of neutrino
production reaching to the FASER$\nu$ detector \cite{Abreu:2019yak}. Moreover, we also consider the subdominant production channel $ D_s \rightarrow l \nu_\alpha Z^\prime$, because of the large mass of $D_s $, $(m_{D_s}= 1968.47\pm0.33 {\rm MeV}) $.
This large mass of strange charm meson is noticeable for large $Z^\prime$ mass since this can allow us to constrain $Z^\prime$ with larger mass range [ $m_{Z^\prime} > 50~ $ {\rm MeV}]. However, there is also a contribution from $D^+$ and $D^-$ decays. We have not taken this into account for simplicity.
The spectrum of neutrinos produced from both meson three-body decay and $Z^\prime$ decay is given by Eq.~\ref{Eq.nuflux}. The main source of the background of electron and tau-neutrinos is the intrinsic background.
Several experiments such as KLOE II \cite{AmelinoCamelia:2010me}, NA48 \cite{Batley:2007aa}, NA62 \cite{Lazzeroni:2012cx}, $K^+ \rightarrow e^+ \nu \bar{\nu} \nu $ \cite{Heintze:1977kk}, $K^+ \rightarrow \mu^+ \nu \bar{\nu} \nu $ \cite{Artamonov:2016wby} and E949 \cite{Zamkovsky:2020obr} have studied kaon decay
with an unprecedented accuracy. For
$ 1~{\rm keV} < m_{Z^\prime} < 2~{\rm MeV}$ also the stringent bound comes from Big Bang Nucleosynthesis (BBN) which is several orders of magnitudes stronger than NA62 bound \cite{Huang:2017egl}. Below 1 {\rm keV}, the most stringent bound is set by NA62.
Moreover, the NA62 experiment \cite{Lazzeroni:2012cx,Bakhti:2017jhm} provides the best measurement and finds the most stringent bound from $1-60 ~{\rm MeV}$. For $ m_{Z^\prime} > 60 ~{\rm MeV}$, the strongest constraint comes from invisible decay of Z \cite{laha}.
We consider pion, kaon and strange charm meson decay and find the constraint on the coupling of a
neutrino with the light gauge boson for $1 ~{\rm KeV} < m_{Z^\prime} < 500 ~{\rm MeV}$. Notice the constraints on the coupling from meson decay has a linear behavior for $ m_{Z^\prime}< 1 ~{\rm MeV}$, thus for $ m_{Z^\prime}< 1 ~{\rm KeV}$ the constrain is a linear extrapolation of the results.
Fig.~\ref{flux_nu_anu} indicates muon, electron and tau (anti-) neutrino fluxes for $m_{Z^\prime}=10~$ {\rm MeV}, assuming standard model and non-zero value for $g_{\alpha\beta}$ equal to 0.001. %We have assumed such a large $g_{e\alpha}$ to demonstrate that the signal is comparable to the background flux.
%Since the most stringent constraint on the parameter space is set by big bang nucleosynthesis \cite{Huang:2017egl}; Thus, we have considered only $ 1 ~{\rm MeV} < m_{Z^\prime} < 50 ~{\rm MeV}$ and cut our figures at 1 MeV.

\begin{figure}[h]

%\hspace{-1cm}

\includegraphics[width=0.7\textwidth]{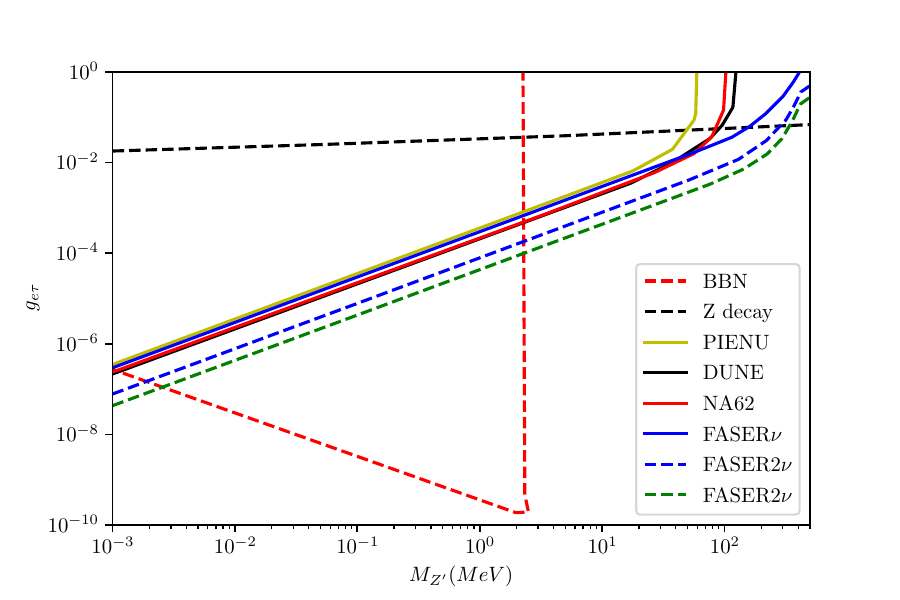}

\caption{The upper bound on $g_{e\tau}$ vs. $m_{Z^\prime}$ at 90\% C.L.
The yellow, red and the blue curves shows the current bound from PIENU \cite{Aguilar-Arevalo:2015cdf}, NA62 \cite{Lazzeroni:2012cx} and FASER$\nu$ \cite{Ariga:2018zuc}, respectively. The black curve corresponds to DUNE ND data assuming ten years of data taking. We have assumed detection efficiency of $2\%$. The blue dashed curve and the green dashed curve indicate the constraints from FASER$2\nu$ corresponding to the assumed data of one hundred times and one thousand times larger than FASER$\nu$, respectively. We have assumed detection efficiency of $80\%$ for FASER$\nu$.
The black dashed line shows the current constraint from Z decay \cite{Laha:2013xua}. The red dashed curve shows the BBN constraint \cite{Huang:2017egl}.}

\label{getau_constraints}

%}

\end{figure}

For statistical inference, we used the chi-squared method. Depending on the number of events in each bin, we used Gaussian or Poisson distribution function, for a large and small number of events, respectively. We have considered the number of events smaller than twenty events as a small number of events. Taking the Asimov data set, we have considered two cases, first the standard model as the true model, and second the new gauge interaction as the true model. We have used the pull method to account for the systematic uncertainties. We have considered the flux normalization uncertainty of $10 \%$. Other important systematic uncertainties come from the shape of the flux of $\nu_e$ and $\nu_{\tau}$. This is very crucial since taking into account the uncertainties in the shape of the background may change the results significantly. Estimating the systematic uncertainties of the $\nu_e$ and $\nu_{\tau}$ spectrum is beyond the scope of this work. During our analysis, we take the couplings to be non-zero
one at a time.
We assume FASER2$\nu$ has $100$ and $1000$ times larger data collection than FASER$\nu$. % and we take the standard model as the true model.
The 90$\%$ constraints on $g_{e\tau}$ are shown in Fig.~\ref{getau_constraints}. The current constraints from meson decay experiments PIENU \cite{Aguilar-Arevalo:2015cdf}, NA62 \cite{Lazzeroni:2012cx} are indicated by yellow and red curves, respectively. Moreover, the current constraint from Z decay \cite{Laha:2013xua} and the BBN constraint \cite{Huang:2017egl} are indicated by the black dashed and red dashed curves. To plot this figure, we assume that $g_{e\tau}$ is non-zero and set the other couplings equal to zero.
Considering 10 years of data taking for DUNE near detector (ND), 5 years in each mode, we have shown the constraint from DUNE ND data by the black curve. The potential of FASER$\nu$ to constrain $g_{e\tau}$ is indicated by the blue curve. We observe that for $ 50 ~{\rm MeV} < m_{Z^\prime} < 100 ~{\rm MeV}$, FASER$\nu$ can constrain $g_{e\tau}$ more stringent than current constraints and DUNE ND. Moreover,
as can be observed for the number of events $100$ (blue dashed) and $1000$ (green dashed) times larger than FASER$\nu$, FASER$2\nu$ can constrain $g_{e\tau}$ stronger than the current constraints for $ m_{Z^\prime} < 2 {\rm keV}$ and $3~{\rm MeV} < m_{Z^\prime} <300 ~{\rm MeV}$.

\begin{figure}[h]

%\hspace{-1cm}

\includegraphics[width=0.7\textwidth]{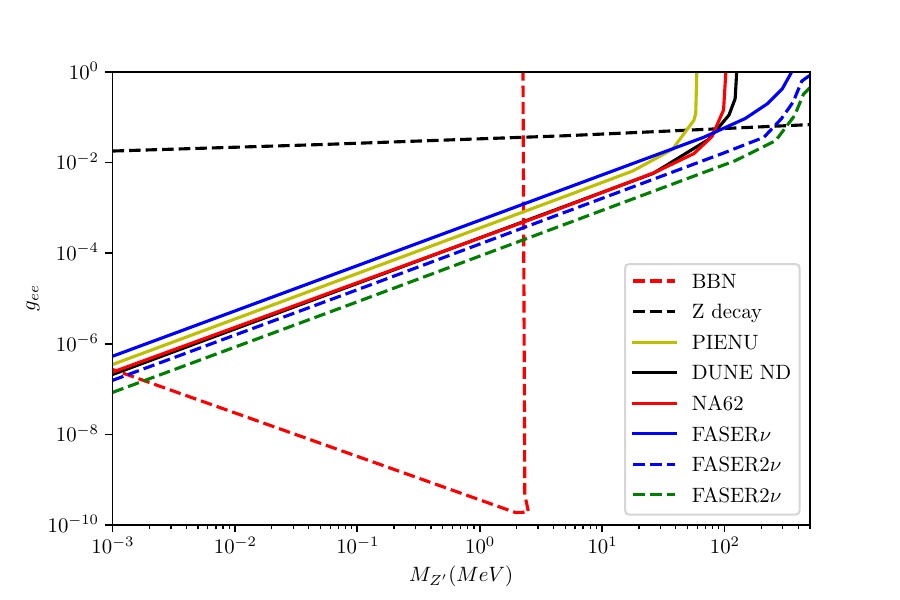}

\caption{The upper bound on $g_{ee}$ vs. $m_{Z^\prime}$ at 90\% C.L.
The yellow, red and the blue curves shows the current bound from PIENU \cite{Aguilar-Arevalo:2015cdf}, NA62 \cite{Lazzeroni:2012cx} and FASER$\nu$ \cite{Ariga:2018zuc}, respectively. The black curve corresponds to DUNE ND data assuming ten years of data taking. The blue dashed curve and the green dashed curve indicate the constraints from FASER$2\nu$ corresponding to the assumed data of one hundred times and one thousand times larger than FASER$\nu$, respectively. We have assumed detection efficiency of $80\%$ for FASER$\nu$.
The black dashed line shows the current constraint from Z decay \cite{Laha:2013xua}. The red dashed curve shows the BBN constraint \cite{Huang:2017egl}.}

\label{gee_constraints}

%}

\end{figure}

It is also interesting to study the case of non-zero $g_{ee}$. In this case, $g_{ee}$ is constrained by detecting electron (anti-)neutrino while in the case of nonzero $g_{e\tau}$, the coupling is constrained from both electron and tau (anti-)neutrino detection.
In Fig.~\ref{gee_constraints} the constraints on $g_{e e}$ are shown. We have assumed that $g_{ee} \neq 0$ while setting other coupling to zero. Although FASER$\nu$ cannot constrain the coupling more stringent than the current one, we observe that FASER$2\nu$ with $100$ and $1000$ times larger than FASER$\nu$ data taking can improve the constraint on $g_{e e}$ for $ m_{Z^\prime} < 2 ~{\rm keV} $ and $3~{\rm MeV} < m_{Z^\prime}< 200 ~{\rm MeV} $.

\begin{figure}[h]

%\hspace{-1cm}

\includegraphics[width=0.7\textwidth]{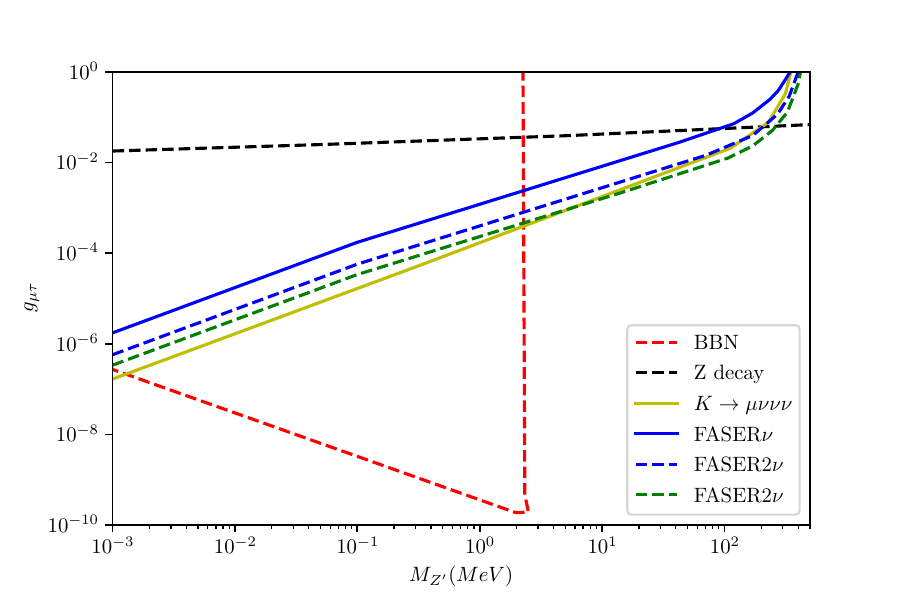}

\caption{The upper bound on $g_{\mu \tau}$ vs. $m_{Z^\prime}$ at 90\% C.L.
The yellow and the blue curves shows the current bound from $K \rightarrow \mu \nu \nu \nu $ \cite{Artamonov:2016wby} and FASER$\nu$, respectively. The blue dashed curve and the green dashed curve indicate the constraints from FASER$2\nu$ corresponding to the assumed data of one hundred times and one thousand times larger than FASER$\nu$, respectively. We have assumed detection efficiency of $80\%$ for FASER$\nu$.
The black dashed line shows the current constraint from Z decay \cite{Laha:2013xua}. The red dashed curve shows the BBN constraint \cite{Huang:2017egl}.}

\label{gmutau_constraints}

%}

\end{figure}

In Fig.~\ref{gmutau_constraints}, we have indicated the upper bound on $g_{\mu \tau}$ vs. $m_{Z^\prime}$ at 90\% C.L., assuming only $g_{\mu \tau} \neq 0$.
The current bound from $K \rightarrow \mu \nu \nu \nu $ \cite{Artamonov:2016wby} and FASER$\nu$ \cite{Ariga:2018zuc} are shown by the yellow and the blue curves, respectively. The black dashed line shows the current constraint from Z decay and the red dashed curve indicates the BBN constraint. As can be observed, for the mass range $10 ~{\rm MeV} < m_{Z^\prime} < 300 ~{\rm MeV} $, FASER$2\nu$ with $1000$ times larger data than FASER$\nu$, can set the most stringent bound on $g_{\mu \tau}$ (green dashed curve).

\begin{figure}[h]
\includegraphics[width=0.7\textwidth]{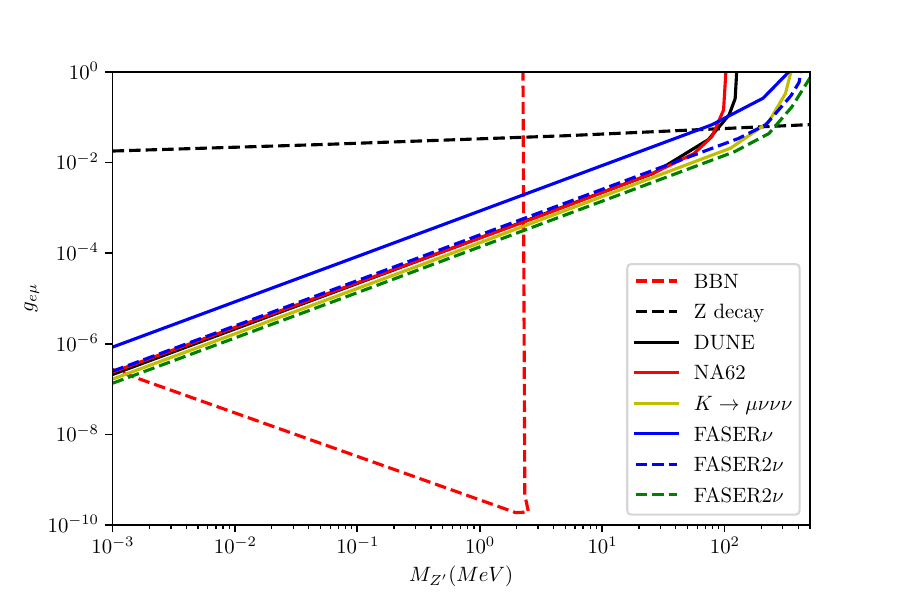}

\caption{The upper bound on $g_{ e \mu }$ vs. $m_{Z^\prime}$ at 90\% C.L.
The yellow, red and the blue curves shows the current bound from $K \rightarrow \mu \nu \nu \nu $ \cite{Artamonov:2016wby}, NA62\cite{Lazzeroni:2012cx} and FASER$\nu$, respectively. The black curve corresponds to DUNE ND data assuming ten years of data taking. The blue dashed curve and the green dashed curve indicate the constraints from FASER$2\nu$ corresponding to the assumed data of one hundred times and one thousand times larger than FASER$\nu$, respectively. We have assumed detection efficiency of $80\%$ for FASER$\nu$.
The black dashed line shows the current constraint from Z decay \cite{Laha:2013xua}. The red dashed curve shows the BBN constraint \cite{Huang:2017egl}.}

\label{gemu_constraints}

%}

\end{figure}

Fig. ~\ref{gemu_constraints} shows $g_{e \mu }$ vs. $m_{Z^\prime}$ at 90\% C.L., assuming only $g_{e \mu } \neq 0$. Again we observe that FASER$2\nu$ with $1000$ times larger data than FASER$\nu$, can improve the constraint slightly in this case, for $15 ~{\rm MeV} < m_{Z^\prime} < 300 ~{\rm MeV} $.

Notice that in the case of $g_{e \tau}$, FASER$\nu$, itself can set bound on $g_{e \tau}$ more stringently than the current constraints as well as DUNE constraint for $50 ~{\rm MeV} < m_{Z^\prime} < 150 ~{\rm MeV} $ while in the case of $g_{e e}$ and $g_{\mu \tau}$ it cannot. This is because FASER$\nu$ can detect $\tau$ neutrinos with high efficiency. Moreover, considering heavy mesons such as strange charm meson which are produced at the interaction point, and their subsequent three-body decay can produce $Z'$ is important to set a constraint on the coupling of neutrino with heavier $Z'$ masses ($50~{\rm MeV}<m_{Z'}<500~{\rm MeV}$).
Our results show that for data $100$ and $1000$ times larger than FASER$\nu$ data, FASER$2\nu$ can improve the constraint on the $g_{ee}$ and $g_{e \tau}$ coupling for $ m_{Z^\prime} < 2 ~{\rm keV} $ and $3~{\rm MeV} < m_{Z^\prime}< 300~{\rm MeV}$. Also FASER$2\nu$ with
$1000$ times larger than FASER$\nu$ data, can slightly improve the constraint on $g_{e \mu}$

%%%%%%%ffff8%%%%%%%%%%%%%%%%%%%%%%%%%%%%%%%%%%%%%%%%%%%%%%%%%%%%%%%%%%%%%

\section{Summary \label{sum}}

We have studied the constraints from meson decay on the coupling of neutrinos to a light new vector boson with a mass smaller than 500~{\rm MeV},
using FASER$\nu$ emulsion detector and its upgraded version, FASER2$\nu$. $Z^\prime $
can be produced via three-body decay of the charged mesons, along with
a charged lepton and a neutrino and can subsequently decay into a neutrino-antineutrino pair before reaching
the detector. The produced neutrinos can be detected at the emulsion detector of FASER$\nu$.

FASER$\nu$, an inexpensive subdetector
of FASER, will provide an opportunity to detect the first collider neutrinos; In particular, FASER$\nu$ will make it possible to study $\nu_e$ and $\nu_\tau$ in detail at the highest energies yet explored.
In this work, we have studied the potential of FASER$\nu$ and proposed an upgraded version of it, FASER2$\nu$, with higher statistics to constrain the secret neutrino gauge interaction. Considering secret neutrino gauge interaction, with $\sum_{\alpha ,\beta} g_{\alpha \beta}Z'_\mu\bar{\nu}_\alpha \gamma^\mu \nu_\beta$ Lagrangian leads to new three-body charged meson decay mode, that charged lepton, neutrino and $Z^\prime$ will be produced and subsequently $Z^\prime$ decays to neutrino antineutrino pair.
% FASER$\nu$ cannot constrain this scenario due to small statistics. FASER2$\nu$ is a proposal of an upgraded version of FASER$\nu$, up to an order of $10^5$ times larger data than FASER$\nu$. We have found that FASER2$\nu$ can constrain the scenario more stringently than the current constraint from NA62 %and the future constraints from the near detector of DUNE
Our results are indicated in Fig.~\ref{getau_constraints} to Fig.~\ref{gemu_constraints}.
As indicated in Fig.~\ref{getau_constraints}, using only
FASER$\nu$ data, for $50 ~{\rm MeV} < m_{Z^\prime} < 150 ~{\rm MeV} $, we can constrain $g_{e \tau}$ more strongly than the current constraints and future DUNE near detector constraint. However we observed that with $100$ and $1000$ times larger than FASER$\nu$, FASER$2\nu$ can improve the limit on $g_{e \tau}$ for $ m_{Z^\prime} < 2 ~{\rm keV} $ and $3~{\rm MeV} < m_{Z^\prime}< 300~{\rm MeV}$.

Moreover, we showed that while FASER$\nu$ is not able to
constrain $g_{ee}$ better than DUNE and the current constraints,
FASER$2\nu$ with a data $100$ and $1000$ times larger than FASER$\nu$ data, can improve the constraint on the $g_{ee}$ for the mass range of $ m_{Z^\prime} < 2 ~{\rm keV} $ and $3~{\rm MeV} < m_{Z^\prime}< 300~{\rm MeV}$ (Fig.~\ref{gee_constraints}).

The results for $g_{\mu \tau}$ was indicated in Fig.~\ref{gmutau_constraints}. We observed that for the mass range
$10 ~{\rm MeV} < m_{Z^{\prime}} < 300 ~{\rm MeV}$, FASER$2\nu$ with 1000 times larger data than FASER$\nu$, can set
the most stringent bound on $g_{\mu \tau}$. Finally we showed that for $g_{e \mu}$, FASER$2\nu$ with 1000 times larger data can just slightly improve the limits (Fig.~\ref{gemu_constraints}) for $ m_{Z^\prime} < 2 ~{\rm keV} $ and $3~{\rm MeV} < m_{Z^\prime}< 300~{\rm MeV}$.

%the total decay rate is
%\begin{align}
%\Gamma(M\longrightarrow e\nu_\alpha Z')&=\frac{ g_{e\alpha}^2 G_F^2 V_{qq^\prime} f_M^2}{6144 \pi ^3 m_M^3% m_{Z'}^2}\left(m_M^8+72 m_M^4 m_{Z'}^4-64 m_M^2 m_{Z'}^6\right.\\
%& \left.+24 \left(3 m_M^4 m_{Z'}^4+4 m_M^2 m_{Z'}^6\right) \log \left(\frac{m_{Z'}}{m_M}\right)-9
%m_{Z'}^8\right)
%\end{align}
%Subsequently, $Z^\prime$ will decay to neutrino anti-neutrino pair with the following decay rate \cite{Bakhti:2018avv}
%\begin{equation}\label{Eq.zpdecay}
%\Gamma(Z'\longrightarrow\nu_\alpha\bar{\nu}_\beta)=\frac{g^2_{\alpha\beta} m_{Z'}}{24\pi}.
%\end{equation}

\subsection*{Acknowledgments}
We are very thankful to the anonymous referees, for the very useful comments and remarks. We are grateful to Y. Farzan for the useful discussion.
This project has received funding from the European Union's Horizon 2020 research and innovation programme under the Marie Sk\l{}odowska-Curie grant agreement No.~674896 and No.~690575. P.B and M.R are grateful to the IFT, UAM
University for warm and generous hospitality. P.B. thanks Iran Science Elites Federation Grant No. 11131. P.B. and M.R. would like to thank the National Research Foundation of Korea Grant (NRF-2020R1I1A3072747).

%%%%%%%%%%%%%%%%%%%%%%%%%%%%%%%%%%%%%%%%%%%%%%%%%%
%%%%%%%%%%%%%%%%%%%%%%%%%%%55KETABNAMEH%%%%%%%%%%%
%%%%%%%%%%%%%%%%%%%%%%%%%%%%%%%%%%%%%%%%%%%%%%%%%%


\begin{thebibliography}{99}

%\cite{Ariga:2018zuc}
\bibitem{Ariga:2018zuc}
A.~Ariga \textit{et al.} [FASER],
%``Letter of Intent for FASER: ForwArd Search ExpeRiment at the LHC,''
[arXiv:1811.10243 [physics.ins-det]].
%31 citations counted in INSPIRE as of 20 Apr 2021
\bibitem{ariga}
FASER Collaboration, A. Ariga et al., “Technical Proposal for FASER: ForwArd Search
ExpeRiment at the LHC,” arXiv:1812.09139 [physics.ins-det].
http://cds.cern.ch/record/2651328. Submitted to the CERN LHCC on 7 November
2018

  %\cite{Abreu:2019yak}
\bibitem{Abreu:2019yak}
H.~Abreu \textit{et al.} [FASER],
%``Detecting and Studying High-Energy Collider Neutrinos with FASER at the LHC,''
Eur. Phys. J. C \textbf{80} (2020) no.1, 61
%doi:10.1140/epjc/s10052-020-7631-5
[arXiv:1908.02310 [hep-ex]].
%15 citations counted in INSPIRE as of 06 Jul 2020

%\cite{Chu:2015ipa}
\bibitem{Chu:2015ipa}
X.~Chu, B.~Dasgupta and J.~Kopp,
%``Sterile neutrinos with secret interactions\textemdash{}lasting friendship with cosmology,''
JCAP \textbf{10} (2015), 011
%doi:10.1088/1475-7516/2015/10/011
[arXiv:1505.02795 [hep-ph]].
%65 citations counted in INSPIRE as of 10 Jan 2021

%\cite{Aarssen:2012fx}
\bibitem{Aarssen:2012fx}
L.~G.~van den Aarssen, T.~Bringmann and C.~Pfrommer,
%``Is dark matter with long-range interactions a solution to all small-scale problems of \textbackslash{}Lambda CDM cosmology?,''
Phys. Rev. Lett. \textbf{109} (2012), 231301
%doi:10.1103/PhysRevLett.109.231301
[arXiv:1205.5809 [astro-ph.CO]].
%192 citations counted in INSPIRE as of 10 Jan 2021



%\cite{Hooper:2007tu}
\bibitem{Hooper:2007tu}
D.~Hooper, M.~Kaplinghat, L.~E.~Strigari and K.~M.~Zurek,
%``MeV Dark Matter and Small Scale Structure,''
Phys. Rev. D \textbf{76} (2007), 103515
%doi:10.1103/PhysRevD.76.103515
[arXiv:0704.2558 [astro-ph]].
%69 citations counted in INSPIRE as of 10 Jan 2021
 
\bibitem{farzan}
Y. Farzan and J. Heeck, Phys. Rev. D 94 (2016) no.5, 053010 [arXiv:1607.07616 [hep-ph]].

\bibitem{farzaan}
Y. Farzan and M. Tortola, Front. in Phys. 6 (2018) 10 [arXiv:1710.09360 [hep-ph]].

\bibitem{Lazzeroni:2012cx}
  C.~Lazzeroni {\it et al.} [NA62 Collaboration],
  %``Precision Measurement of the Ratio of the Charged Kaon Leptonic Decay Rates,''
  Phys.\ Lett.\ B {\bf 719} (2013) 326
  %doi:10.1016/j.physletb.2013.01.037
  [arXiv:1212.4012 [hep-ex]].
  %%CITATION = doi:10.1016/j.physletb.2013.01.037;%%
  %114 citations counted in INSPIRE as of 04 Sep 2019








\bibitem{Bakhti:2017jhm}
  P.~Bakhti and Y.~Farzan,
  %``Constraining secret gauge interactions of neutrinos by meson decays,''
  Phys.\ Rev.\ D {\bf 95} (2017) no.9,  095008
 % doi:10.1103/PhysRevD.95.095008
  [arXiv:1702.04187 [hep-ph]].
  %%CITATION = doi:10.1103/PhysRevD.95.095008;%%
  %13 citations counted in INSPIRE as of 03 Sep 2019

\bibitem{Acciarri:2015uup}
  R.~Acciarri {\it et al.} [DUNE Collaboration],
  %``Long-Baseline Neutrino Facility (LBNF) and Deep Underground Neutrino Experiment (DUNE) : Conceptual Design Report, Volume 2: The Physics Program for DUNE at LBNF,''
  arXiv:1512.06148 [physics.ins-det].
  %%CITATION = ARXIV:1512.06148;%%
  %496 citations counted in INSPIRE as of 04 Sep 2019



  \bibitem{Bakhti:2018avv}
  P.~Bakhti, Y.~Farzan and M.~Rajaee,
  %``Secret interactions of neutrinos with light gauge boson at the DUNE near detector,''
  Phys.\ Rev.\ D {\bf 99} (2019) no.5,  055019
 % doi:10.1103/PhysRevD.99.055019
  [arXiv:1810.04441 [hep-ph]].
  %%CITATION = doi:10.1103/PhysRevD.99.055019;%%
  %5 citations counted in INSPIRE as of 03 Sep 2019

\bibitem{Kling:2020iar}
F.~Kling,
%``Probing light gauge bosons in tau neutrino experiments,''
Phys. Rev. D \textbf{102} (2020) no.1, 015007
%doi:10.1103/PhysRevD.102.015007
[arXiv:2005.03594 [hep-ph]].
%11 citations counted in INSPIRE as of 23 Apr 2021

\bibitem{Farzan:2016wym}
Y.~Farzan and J.~Heeck,
%``Neutrinophilic nonstandard interactions,''
Phys. Rev. D \textbf{94} (2016) no.5, 053010
%doi:10.1103/PhysRevD.94.053010
[arXiv:1607.07616 [hep-ph]].
%54 citations counted in INSPIRE as of 23 Dec 2020


%\cite{Fernandez-Martinez:2016lgt}
\bibitem{Fernandez-Martinez:2016lgt}
E.~Fernandez-Martinez, J.~Hernandez-Garcia and J.~Lopez-Pavon,
%``Global constraints on heavy neutrino mixing,''
JHEP \textbf{08} (2016), 033
%doi:10.1007/JHEP08(2016)033
[arXiv:1605.08774 [hep-ph]].
%158 citations counted in INSPIRE as of 23 Dec 2020



\bibitem{Ball:2017nwa}
  R.~D.~Ball {\it et al.} [NNPDF Collaboration],
  %``Parton distributions from high-precision collider data,''
  Eur.\ Phys.\ J.\ C {\bf 77} (2017) no.10,  663
  %doi:10.1140/epjc/s10052-017-5199-5



%\cite{AmelinoCamelia:2010me}
\bibitem{AmelinoCamelia:2010me}
G.~Amelino-Camelia, F.~Archilli, D.~Babusci, D.~Badoni, G.~Bencivenni, J.~Bernabeu, R.~A.~Bertlmann, D.~R.~Boito, C.~Bini and C.~Bloise, \textit{et al.}
%``Physics with the KLOE-2 experiment at the upgraded DA$\phi$NE,''
Eur. Phys. J. C \textbf{68} (2010), 619-681
%doi:10.1140/epjc/s10052-010-1351-1
[arXiv:1003.3868 [hep-ex]].
%335 citations counted in INSPIRE as of 10 Jan 2021

%\cite{Batley:2007aa}
\bibitem{Batley:2007aa}
J.~R.~Batley \textit{et al.} [NA48/2],
%``Search for direct CP violating charge asymmetries in K+- ---\ensuremath{>} pi+- pi+ pi- and K+- ---\ensuremath{>} pi+- pi0 pi0 decays,''
Eur. Phys. J. C \textbf{52} (2007), 875-891
%doi:10.1140/epjc/s10052-007-0456-7
[arXiv:0707.0697 [hep-ex]].
%169 citations counted in INSPIRE as of 10 Jan 2021






%\cite{Heintze:1977kk}
\bibitem{Heintze:1977kk}
J.~Heintze, G.~Heinzelmann, P.~Igo-Kemenes, R.~Mundhenke, H.~Rieseberg, B.~Schurlein, H.~W.~Siebert, V.~Soergel, H.~Stelzer and K.~P.~Streit, \textit{et al.}
%``A Measurement of the K+ --\ensuremath{>} e+ Neutrino gamma Structure Decay,''
Nucl. Phys. B \textbf{149} (1979), 365-380
%doi:10.1016/0550-3213(79)90001-4
%59 citations counted in INSPIRE as of 12 Jan 2021

\bibitem{Artamonov:2016wby}
A.~V.~Artamonov \textit{et al.} [E949],
%``Search for the rare decay $K^+\to\mu^+\nu\bar\nu\nu$,''
Phys. Rev. D \textbf{94} (2016) no.3, 032012
%doi:10.1103/PhysRevD.94.032012
[arXiv:1606.09054 [hep-ex]].
%11 citations counted in INSPIRE as of 23 Apr 2021

%\cite{Zamkovsky:2020obr}
\bibitem{Zamkovsky:2020obr}
M.~Zamkovsk\'y,
%``Study of the extremely rare decay $K^+ \to \pi^+\nu \bar\nu$ with the NA62 experiment at CERN,''
%0 citations counted in INSPIRE as of 10 Jan 2021


\bibitem{Huang:2017egl}
G.~y.~Huang, T.~Ohlsson and S.~Zhou,
%``Observational Constraints on Secret Neutrino Interactions from Big Bang Nucleosynthesis,''
Phys. Rev. D \textbf{97} (2018) no.7, 075009
%doi:10.1103/PhysRevD.97.075009
[arXiv:1712.04792 [hep-ph]].
%38 citations counted in INSPIRE as of 21 Aug 2020,


\bibitem{laha}
 R. Laha, B. Dasgupta and J. F. Beacom, 
%“Constraints on new neutrino interactions via light Abelian vector bosons,” 
Phys. Rev. D 89, 093025 (2014) [arXiv:1304.3460].

\bibitem{Bordag:2001qi}
  M.~Bordag, U.~Mohideen and V.~M.~Mostepanenko,
  %``New developments in the Casimir effect,''
  Phys.\ Rept.\  {\bf 353} (2001) 1
 % doi:10.1016/S0370-1573(01)00015-1
  [quant-ph/0106045].
  %%CITATION = doi:10.1016/S0370-1573(01)00015-1;%%
  %839 citations counted in INSPIRE as of 06 Sep 2019



 %\cite{Kozhuharov:2014dji}
\bibitem{Kozhuharov:2014dji}
V.~Kozhuharov [NA62],
%``Measurement of the ratio of the charged kaon leptonic decays at NA62,''
Int. J. Mod. Phys. Conf. Ser. \textbf{35} (2014), 1460436
%doi:10.1142/S2010194514604360
[arXiv:1412.0243 [hep-ex]].
%0 citations counted in INSPIRE as of 21 Jul 2020





%\cite{Aguilar-Arevalo:2015cdf}
\bibitem{Aguilar-Arevalo:2015cdf}
A.~Aguilar-Arevalo \textit{et al.} [PiENu],
%``Improved Measurement of the $\pi \to \textrm{e} \nu$ Branching Ratio,''
Phys. Rev. Lett. \textbf{115} (2015) no.7, 071801
%doi:10.1103/PhysRevLett.115.071801
[arXiv:1506.05845 [hep-ex]].
%59 citations counted in INSPIRE as of 20 Apr 2021




%\cite{Huang:2017egl}
\bibitem{Huang:2017egl}
G.~y.~Huang, T.~Ohlsson and S.~Zhou,
%``Observational Constraints on Secret Neutrino Interactions from Big Bang Nucleosynthesis,''
Phys. Rev. D \textbf{97} (2018) no.7, 075009
%doi:10.1103/PhysRevD.97.075009
[arXiv:1712.04792 [hep-ph]].
%45 citations counted in INSPIRE as of 20 Apr 2021



%\cite{Laha:2013xua}
\bibitem{Laha:2013xua}
R.~Laha, B.~Dasgupta and J.~F.~Beacom,
%``Constraints on New Neutrino Interactions via Light Abelian Vector Bosons,''
Phys. Rev. D \textbf{89} (2014) no.9, 093025
%doi:10.1103/PhysRevD.89.093025
[arXiv:1304.3460 [hep-ph]].
%76 citations counted in INSPIRE as of 20 Apr 2021



\bibitem{Abreu:2020ddv}
H.~Abreu \textit{et al.} [FASER],
%``Technical Proposal: FASERnu,''
[arXiv:2001.03073 [physics.ins-det]].
%25 citations counted in INSPIRE as of 24 May 2021






  























  \end{thebibliography}
\end{document}